# Enhancing Mobile Ad Hoc Networks (MANETs) with Software-Defined Networking (SDN): A Balanced Approach

Riccardo Fonti and Andrea Piroddi*

Department of Engineering and Computer Science, Faculty of Engineering, University of Bologna, Cesena, Italy
Email: Riccardo.fonti@studio.unibo.it (R.F.); andrea.piroddi@unibo.it (A.P.)
*Corresponding author

*Abstract*—Mobile Ad Hoc Networks (MANETs) are decentralized wireless networks, characterized by their dynamic topologies and node mobility. In the era of cutting-edge technologies, integrating Software-Defined Networking (SDN) with MANETs offers a promising solution to manage these challenges more efficiently. This paper presents a balanced discussion of MANETs and SDN, demonstrating how SDN principles, such as centralized control and network virtualization, can optimize MANET performance in terms of scalability, cost-efficiency, and security. A mathematical model is developed to analyze Capital Expenditures (CAPEX), Operational Expenditures (OPEX), and network efficiency.

## I. INTRODUCTION

Mobile Ad Hoc Networks (MANETs) are a type of wireless network that are dynamically formed by mobile devices without the need for any pre-existing infrastructure, such as routers or access points [1]. These networks are self-configuring and decentralized, meaning that each device in the network can act both as a host and a router, forwarding data to other devices. MANETs are highly adaptive to changes in network topology, which makes them suitable for environments where network infrastructure is either not feasible or too expensive to deploy, such as disaster recovery scenarios, military operations, and sensor networks [2].

However, the decentralized nature of MANETs poses significant challenges, particularly in areas such as network control, resource management, and security. The lack of centralized control often leads to inefficient routing, limited scalability, and vulnerabilities to attacks such as spoofing or Denial of Service (DoS) [3]. Moreover, due to the dynamic nature of MANETs, nodes frequently change their locations, making it difficult to maintain stable connections and effectively manage resources, such as bandwidth and power consumption [4]. This necessitates the need for efficient mechanisms that can handle these complexities while maintaining network performance.

In contrast, Software-Defined Networking (SDN) introduces a paradigm where the control plane is decoupled from the data plane, allowing for centralized network control and management [5]. SDN has gained attention in recent years for its potential to address many of the challenges associated with traditional network architectures by offering flexibility, programmability, and centralized management. By integrating SDN principles into MANETs, the network's inherent challenges, such as dynamic topology changes, resource constraints, and security vulnerabilities, can be mitigated. SDN's centralized control plane can provide a global view of the network, enabling more efficient routing, better resource allocation, and improved security [6].

The integration of SDN and MANETs represents a promising direction for improving the scalability, flexibility, and cost-efficiency of these networks [7]. Centralized control through SDN allows for real-time monitoring and decision-making, which can enhance MANET performance in terms of bandwidth utilization and power efficiency. Moreover, SDN's programmability allows for dynamic reconfiguration of network policies, leading to more robust and adaptable networks in unpredictable environments [5].

This paper explores how SDN principles can enhance MANET architectures, specifically addressing the issues of scalability, resource management, and security. Through a mathematical model we quantify the potential cost and efficiency improvements brought by the integration of SDN in MANETs.

## II. MANET CHARACTERISTICS AND CHALLENGES

Mobile Ad Hoc Networks (MANETs) are unique in their ability to establish wireless networks on-the-fly, without the need for fixed infrastructure. This capability makes them highly suitable for scenarios like military operations, disaster recovery, and temporary event coverage [1]. However, the same characteristics that give MANETs their flexibility also introduce several technical challenges, especially regarding network control, scalability, and resource management.

### A. Dynamic Topology

One of the defining features of MANETs is their dynamic topology. The mobility of the nodes results in frequent changes to network routes as connections are established and broken unpredictably [3]. Each node in a MANET functions both as a host and a router, forwarding packets for other nodes, which means that routing protocols must continuously adapt to the changing topology [2]. Traditional routing protocols like Ad-hoc On-Demand Distance Vector (AODV) and Dynamic Source Routing (DSR) have been developed to address these challenges, but they face limitations in large-scale and highly dynamic environments. These protocols rely heavily on route discovery and maintenance processes, which can incur significant overhead as the network scales or becomes more dynamic.

Moreover, the lack of a fixed infrastructure means that routing decisions must be made in a distributed manner.

This decentralized control often leads to suboptimal routing paths, increased latency, and packet losses during route recalculations. Additionally, link breakages due to node mobility can severely degrade the performance of the network, particularly when nodes move rapidly, causing frequent route disruptions [4].

*B. Scalability*

Scalability is another critical issue in MANETs. As the number of nodes increases, the complexity of maintaining up-to-date routing tables and managing network resources grows exponentially [3]. In a large MANET, nodes are required to store and process a greater amount of routing information, leading to increased computational overhead and higher power consumption. Furthermore, broadcast storms—where nodes flood the network with control messages—can become a significant problem as the network size increases [8].

In scenarios where the density of nodes is high, the probability of interference between nodes also rises, which further limits the scalability of the network. Moreover, the dynamic nature of the topology requires routing protocols to constantly update routes, which results in increased signaling traffic that can congest the network, limiting its scalability. Solutions such as hierarchical routing, zone-based routing, and clustering have been proposed to address scalability issues, but these approaches introduce additional complexity in managing the network [9].

*C. Resource Constraints*

Devices in MANETs typically operate in resource-constrained environments. Nodes are often equipped with limited battery power, processing capabilities, and bandwidth [4]. Power consumption is a particularly critical issue, as mobile devices need to conserve energy to extend their operational lifetime. The overhead associated with constant route updates, packet forwarding, and participation in routing processes can quickly drain battery life, especially in dense or highly mobile networks.

Bandwidth constraints also pose significant challenges. In MANETs, wireless links are shared among multiple nodes, leading to contention for the available bandwidth. This contention can result in lower data throughput, higher packet collision rates, and increased delays [10, 11]. Furthermore, the dynamic topology exacerbates these issues, as frequent route changes and packet retransmissions consume additional bandwidth. Effective bandwidth management, such as through load balancing and congestion control mechanisms, is crucial for ensuring optimal network performance.

Addressing these challenges is essential for improving the reliability and performance of MANETs. Solutions such as cross-layer optimization, energy-efficient routing protocols, and adaptive resource management have been proposed to tackle these issues, but they are often limited in scope or introduce trade-offs in terms of complexity and performance [9].

## III. SDN IN MANETs: INTEGRATION FRAMEWORK

Software-Defined Networking (SDN) introduces a centralized approach to network control, which contrasts the distributed nature of traditional MANETs. By decoupling the control plane from the data plane, SDN provides centralized decision-making and control over the entire network. This separation allows SDN controllers to manage the routing, resource allocation, and security policies of MANETs, improving overall network performance [5].

*A. Centralized Control*

In traditional MANETs, routing decisions are made locally by each node using distributed algorithms like AODV (Ad-hoc On-Demand Distance Vector) or OLSR (Optimized Link State Routing). These protocols can be inefficient in highly dynamic environments due to frequent route changes and the overhead of route discovery. SDN introduces a global, centralized controller that manages routing across the network with a full view of the topology.

Let us consider a MANET consisting of $n$ nodes, each with a degree of connectivity denoted as $d_i$, where $i = 1, 2, ..., n$. In a traditional distributed system, the routing complexity increases proportionally with the degree $d_i$, as each node maintains routes to its neighbors independently [3]. The average path cost $C_{\text{path}}$ in a decentralized MANET can be represented as:

$$C_{\text{path}} = \frac{1}{n}\sum_{i=1}^{n} C_i$$

where $C_i$ represents the path cost for node $i$. As node mobility increases, $C_i$ grows due to the need for frequent route recalculations and the overhead of broadcasting route updates.

With SDN, the controller computes and optimizes routes centrally, minimizing redundant route discovery and reducing the overall path cost. The centralized routing decision can be modeled as:

$$C_{\text{SDN-path}} = \min_{P} \sum_{i=1}^{n} w_i \cdot d_i$$

where $w_i$ is a weight associated with each node's traffic load or priority, and $P$ represents all possible paths. The SDN controller can dynamically adjust routes based on real-time traffic conditions, leading to lower latency and packet loss, as shown by studies on OpenFlow-based SDN implementations [5].

*B. Dynamic Resource Allocation*

One of the major benefits of SDN in MANETs is dynamic resource allocation. In a resource-constrained environment like MANETs, where nodes have limited power, bandwidth, and computational capacity, efficient resource management is critical. Traditional MANET protocols lack the capability to manage resources dynamically, as each node independently handles its power and bandwidth allocation, often leading to suboptimal usage and network congestion [4].

With SDN, the controller can monitor the network state and allocate resources dynamically based on the real-time demands of the network. Let $B_{\text{total}}$ be the total available bandwidth, and $P_{\text{total}}$ be the total available power in the network. The SDN controller aims to maximize the resource utilization by minimizing the allocation costs $C_{\text{alloc}}$, which can be defined as:

$$C_{\text{alloc}} = \sum_{i=1}^{n} P(V_i) \cdot I(V_i)$$

where $B_i$ and $P_i$ represent the bandwidth and power allocated to node $i$, respectively. The controller uses this cost function to optimize the allocation of resources, ensuring that no node is overburdened or starved of resources [7].

Additionally, SDN can implement load-balancing techniques, distributing traffic across multiple paths to avoid congestion. By dynamically adjusting resource allocation, the controller can reduce the likelihood of bottlenecks and ensure more equitable distribution of resources across the network.

### C. Improved Security

Security is another critical challenge in MANETs, especially due to their decentralized nature and the absence of a trusted central authority. Traditional MANETs are vulnerable to various attacks, such as spoofing, blackhole, and denial-of-service (DoS) attacks. In decentralized networks, each node must independently implement security protocols, which often leads to inconsistent enforcement and gaps in the network's defense [9].

SDN offers a centralized security framework where the controller can enforce uniform security policies across the entire network. The controller can monitor traffic in real-time and detect anomalies or suspicious activity, such as a sudden spike in traffic (indicating a potential DoS attack) or an unauthorized node attempting to join the network. Let $S_{\text{risk}}$ represent the security risk in a MANET, which is a function of the number of vulnerabilities $V$ and the impact of those vulnerabilities $I$:

$$S_{\text{risk}} = \sum_{i=1}^{n} \left( \frac{B_i}{B_{total}} + \frac{P_i}{P_{total}} \right)$$

where $P(V_i)$ is the probability of vulnerability $V_i$ being exploited and $I(V_i)$ is the potential impact of such an exploit.

With SDN, the controller can dynamically adjust security policies, such as rerouting traffic away from compromised nodes or isolating malicious nodes from the network. This reduces both the probability of successful attacks and their potential impact, significantly improving the overall security of the network [5].

### D. SDN Protocols in MANETs

Protocols like OpenFlow are central to the SDN framework, as they enable secure communication between the control plane and the network devices [5]. In the context of MANETs, integrating OpenFlow helps manage the complexity of routing in a dynamic environment by providing more efficient routing paths and reducing the overhead associated with route discovery processes.

The OpenFlow protocol allows SDN controllers to update routing tables in real-time, enabling the network to adapt quickly to changes in topology. Let $T_{\text{update}}$ represent the time it takes to update the routing table in a traditional MANET, which can be expressed as:

$$T_{\text{update}} = T_{\text{discovery}} + T_{\text{propagation}} + T_{\text{reconfiguration}}$$

where $T_{\text{discovery}}$ is the time taken to discover a new route, $T_{\text{propagation}}$ is the time for routing updates to propagate through the network, and $T_{\text{reconfiguration}}$ is the time to reconfigure affected nodes. In an SDN-based MANET, the controller reduces this update time by centrally managing the routing process, minimizing $T_{\text{discovery}}$ and $T_{\text{propagation}}$, as the controller has a global view of the network. This leads to faster route reconfigurations, improving the overall responsiveness of the network [7].

By leveraging SDN protocols, MANETs can benefit from more flexible and programmable network configurations, allowing for better resource management and faster adaptation to changing conditions. This integration opens up new possibilities for optimizing the performance of MANETs, particularly in environments where the network is subject to rapid and unpredictable changes.

Integrating OpenFlow into MANETs allows for more efficient routing and network management, especially in dynamic environments where network topology constantly changes.

In a traditional MANET, nodes must rely on distributed routing protocols such as AODV or OLSR to discover and maintain routes. These protocols, however, can become inefficient as network size increases or when nodes frequently move, requiring constant route recalculation. OpenFlow mitigates these issues by introducing a centralized SDN controller that maintains a global view of the network topology and can update routing rules dynamically.

- Example: OpenFlow-based Route Optimization in MANETs

Consider a MANET consisting of five mobile nodes: *A*, *B*, *C*, *D*, and *E*, where nodes *A* and *E* need to communicate. In a traditional MANET, routing would rely on a reactive protocol like AODV, where node *A* initiates a route discovery process, broadcasting a route request (RREQ) to its neighbors. Nodes *B*, *C*, and *D* forward the RREQ until it reaches node *E*. This process generates considerable overhead, especially in dense or large networks, where multiple nodes may be involved in the route discovery process.

With OpenFlow integrated into an SDN-based MANET, the scenario would unfold differently. First, the SDN controller is deployed, maintaining a global view of the network's current topology. When node *A* wants to send data to node *E*, it sends a request to the SDN controller, which calculates the optimal path based on network conditions such as node mobility, traffic load, and link quality. The controller can then configure OpenFlow-enabled switches or nodes to establish the route.

- Mathematical Representation of Route Optimization:

Let the network be represented as a graph $G = (V, E)$, where $V$ is the set of nodes (mobile devices) and $E$ is the set of wireless links between nodes. The SDN controller computes the shortest or optimal path $P$ from node $A$ to node $E$ by solving the following optimization problem:

$$P_{A \to E} = \arg\min_{P} \sum_{(i,j) \in P} w_{ij}$$

where $w_{ij}$ represents the weight of the link between nodes $i$ and $j$, which could be a function of link quality, delay, or bandwidth availability. The controller uses real-time data from the network to assign weights dynamically.

Once the optimal path $P_{A \to E}$ is computed, the controller sends flow rules to the OpenFlow-enabled devices along the path, installing routing rules that forward packets from node *A* to node *E* through the intermediate nodes (e.g., *B* and *C*). The use of flow rules significantly reduces the overhead associated with route discovery and maintenance, as routes are established proactively and optimized based on real-time network conditions [6].

- Flow Table Example:

An OpenFlow-enabled node in the network maintains a flow table that stores the routing rules for different data flows. For example, the flow table at node *B* might look like this in Table 1:

Table 1. Flow table example

| Match | Action | Priority |
|---|---|---|
| Source: *A*, Destination: *E* | Forward to *C* | High |
| Source: *D*, Destination: *F* | Forward to *E* | Medium |

The match field indicates the source and destination of the packets, while the action field specifies the next hop in the forwarding process. This setup allows the network to forward packets along the optimal path with minimal delay, as nodes no longer need to independently compute routes.

- Dynamic Reconfiguration:

One of the key advantages of using OpenFlow in SDN-enabled MANETs is dynamic reconfiguration. In a traditional MANET, if the link between nodes *B* and *C* were to break due to node movement, the network would need to initiate a new route discovery process, leading to increased delay and packet loss. In an OpenFlow-based MANET, the SDN controller can detect the broken link and immediately compute an alternative route (e.g., $A \to D \to E$) and update the flow tables of the affected nodes.

This ability to dynamically reconfigure routes on the fly reduces the time-to-repair and improves the overall reliability of the network. The total reconfiguration time $T_{\text{reconfig}}$ is given by:

$$T_{\text{reconfig}} = T_{\text{detection}} + T_{\text{computation}} + T_{\text{update}}$$

where:
- $T_{\text{detection}}$ is the time it takes for the SDN controller to detect the link failure.
- $T_{\text{computation}}$ is the time taken to compute a new route.
- $T_{\text{update}}$ is the time required to update the flow tables in the affected nodes.

Because the SDN controller has a global view of the network, $T_{\text{computation}}$ is significantly lower than in traditional distributed systems, where each node would need to independently discover the new route. This results in faster recovery from link failures and improved network resilience [5].

*E. Advantages of Using OpenFlow in MANETs*

The integration of OpenFlow into MANETs offers several key advantages:
- **Improved Routing Efficiency:** With the controller managing routes globally, OpenFlow reduces the overhead associated with distributed route discovery.
- **Real-time Network Adaptation:** The SDN controller continuously monitors network conditions, allowing for real-time adjustments to routing paths.
- **Enhanced Security:** OpenFlow enables centralized security management, allowing the SDN controller to apply uniform security policies across the network.
- **Reduced Latency:** By eliminating the need for reactive route discovery, OpenFlow can reduce latency and packet loss in highly dynamic environments.

The example presented above illustrates how OpenFlow can optimize route computation and dynamically reconfigure the network to respond to changes in topology. By leveraging SDN protocols, MANETs can achieve more efficient, secure, and adaptable network management, making them suitable for a wide range of applications, including military operations, disaster recovery, and IoT environments.

IV. MATHEMATICAL MODELING OF COST AND EFFICIENCY

The economic benefits of integrating SDN with MANETs can be analyzed using mathematical models that quantify both capital expenditures (CAPEX) and operational expenditures (OPEX). In this section, we compare the CAPEX and OPEX models for traditional MANETs and SDN-enabled MANETs, illustrating the cost advantages of the SDN-MANET approach.

*A. CAPEX Model*

Traditional MANETs require specialized hardware, such as routers and switches, capable of independently handling routing and management tasks at each node. This decentralized architecture means that each node must have sufficient computational power and specialized hardware to perform routing, security, and other network management functions [4]. The CAPEX for traditional MANETs is heavily dependent on the number of nodes n and the cost of hardware components at each node.

For a traditional MANET, the total CAPEX is modeled as:

$$CAPEX_{\text{MANET}} = \sum_{i=1}^{n} \text{Cost}_{hwi} + \sum_{i=1}^{n} \text{Cost}_{swi}$$

where:
- $\text{Cost}_{hw_i}$ is the cost of the hardware at node *i*, which includes specialized routers, switches, and other devices.
- $\text{Cost}_{sw_i}$ is the cost of software at node *i*, required for routing, security, and network management.

In SDN-enabled MANETs, the architecture is centralized in the control plane. This means that the routing and network management decisions are made centrally by the SDN controller, reducing the need for specialized hardware at each node. Instead, nodes in SDN-MANETs can rely on general-purpose hardware, with the SDN controller handling the complex computational tasks.

The CAPEX for SDN-enabled MANETs is modeled as:

$$CAPEX_{\text{SDN}} = \sum_{i=1}^{n} \text{Cost}_{hwi} + \text{Cost}_{controller}$$

where:

- $\text{Cost}_{hw_i}$ is the cost of general-purpose hardware at node $i$.
- $\text{Cost}_{controller}$ is the cost of the centralized SDN controller.

Comparison: The key difference in CAPEX between traditional MANETs and SDN-enabled MANETs is that in SDN-MANETs, the reliance on general-purpose hardware reduces the overall cost per node. While there is an additional cost for the SDN controller, the total CAPEX is reduced because each node no longer needs to perform complex routing and management tasks, reducing the need for specialized hardware. As the number of nodes increases, the cost advantage of SDN-MANETs becomes more pronounced, as the per-node hardware cost in SDN-MANETs is lower.

### B. OPEX Model

In traditional MANETs, the Operational Expenditures (OPEX) are mainly driven by maintenance, configuration, and monitoring costs at each node. Since each node is responsible for independently handling routing and management tasks, the cost of maintaining and updating software and hardware at each node is high [3]. The total OPEX for traditional MANETs is expressed as:

$$\text{OPEX}_{\text{MANET}} = \sum_{i=1}^{n}(C_{\text{manutMANET}_i} + C_{\text{configMANET}_i} + C_{\text{monitoraggioMANET}_i})$$

where:
- $C_{\text{manutMANET}_i}$ is the maintenance cost at node $i$.
- $C_{\text{configMANET}_i}$ is the configuration cost at node $i$.
- $C_{\text{monitoraggioMANET}_i}$ is the monitoring cost at node $i$.

In SDN-enabled MANETs, the centralized control plane simplifies network management, reducing the operational costs associated with configuring and maintaining each node. The SDN controller handles most of the configuration and monitoring tasks, leading to lower OPEX at the node level. The OPEX for SDN-enabled MANETs is expressed as:

$$\text{OPEX}_{SDN} = C_{\text{manutSDN}} + C_{\text{configSDN}} + C_{\text{monitoraggioSDN}} + \sum_{i=1}^{n} C_{\text{manutNode}_i}$$

where:
- $C_{\text{manutSDN}}$ is the maintenance cost for the SDN controller.
- $C_{\text{configSDN}}$ is the configuration cost for the SDN controller.
- $C_{\text{monitoraggioSDN}}$ is the monitoring cost handled by the SDN controller.
- $C_{\text{manutNode}_i}$ is the reduced maintenance cost for node $i$, as it no longer handles complex routing tasks.

Comparison: In traditional MANETs, each node incurs significant maintenance, configuration, and monitoring costs. In contrast, SDN-enabled MANETs centralize these tasks in the SDN controller, reducing the operational burden on individual nodes. This results in lower overall OPEX in SDN-MANETs, particularly in large networks where the number of nodes is high.

### C. Efficiency Gains

The integration of SDN into MANETs not only reduces CAPEX and OPEX but also significantly improves network efficiency. In traditional MANETs, routing decisions are made locally at each node, which can lead to suboptimal routes and increased network congestion [5]. SDN, with its global view of the network, can optimize routing and dynamically balance network loads, improving efficiency.

The efficiency of SDN-enabled MANETs can be expressed as:

$$\eta_{\text{SDN}} = \frac{\text{Useful Data}}{\text{Total Bandwidth}} \times \eta_{\text{optimization}}$$

where:
- $\frac{\text{Useful Data}}{\text{Total Bandwidth}}$ is the efficiency of data transmission.
- $\eta_{\text{optimization}} > 1$ reflects the increase in efficiency due to SDN's ability to optimize traffic dynamically.

In traditional MANETs, the absence of centralized control can lead to inefficient bandwidth usage, as nodes must rely on local information for routing. In contrast, SDN-enabled MANETs can optimize bandwidth allocation and ensure that data is transmitted along the most efficient paths, reducing delays and increasing throughput.

## V. SCALABILITY AND SECURITY IN SDN-MANETs

SDN enhances the scalability of MANETs by allowing dynamic management of network resources, which is crucial for handling an increasing number of nodes. The centralized controller in SDN-MANETs can efficiently manage network resources, ensuring that the network remains scalable as the number of nodes increases. Security is also improved, as SDN enables centralized monitoring and real-time response to threats, reducing the risk of attacks such as Denial of Service (DoS) and spoofing.

### A. Scalability Model

Scalability is one of the most critical aspects of MANETs, especially in environments where the number of nodes and their mobility are subject to rapid changes. Traditional MANETs face significant challenges in maintaining scalability due to the decentralized nature of network management. As the network grows in size, maintaining efficient communication paths and ensuring optimal resource usage becomes increasingly difficult [4]. Each node in a traditional MANET is responsible for routing and management, which increases the overhead in terms of computational power, bandwidth consumption, and network congestion.

In contrast, SDN-enabled MANETs introduce a centralized control plane via the SDN controller, which significantly enhances the network's scalability. The SDN controller has a global view of the network and can dynamically manage resources, optimize routing paths, and allocate bandwidth in real-time. The scalability of an SDN-enabled MANET can be modeled by considering both the individual node capacities and the additional scalability introduced by the SDN controller.

$$\text{Capacity}_{\text{SDN}} = \sum_{i=1}^{n} \text{Capacity}_i + \text{Capacity}_{SDNController}$$

where:
- $\text{Capacity}_i$ represents the capacity of each individual node $i$, which includes computational power,

bandwidth, and routing capabilities.
- Capacity$_{SDNController}$ is the additional capacity provided by the SDN controller, which manages the global routing and resource allocation tasks.

*B. Scalability Advantages of SDN-Enabled MANETs*

- **Centralized Resource Management:** The SDN controller manages resources across the network, optimizing the allocation of bandwidth and computational power, thus preventing network congestion and ensuring load balancing [5]. This centralized management prevents bottlenecks in dense networks and reduces the chance of overloading individual nodes.
- **Efficient Routing:** Traditional MANETs often suffer from routing inefficiencies due to the lack of a global view of the network. Nodes make routing decisions based on local information, leading to suboptimal paths, increased packet loss, and higher latencies. In SDN-enabled MANETs, the controller dynamically computes optimal routes for all nodes, minimizing the number of hops and ensuring more efficient routing paths [6]. This is particularly beneficial as the number of nodes increases.
- **Handling Node Mobility:** MANETs are characterized by the high mobility of nodes, which frequently leads to broken links and route recalculations. In traditional MANETs, route discovery is initiated locally by each node, resulting in increased control message overhead and latency. In SDN-enabled MANETs, the controller can preemptively manage link failures by quickly recalculating routes and updating flow tables, significantly improving network responsiveness and maintaining performance even in highly dynamic environments [7].
- **Network Partitioning and Clustering:** As the number of nodes grows, SDN-enabled MANETs can implement hierarchical or cluster-based management. The SDN controller can create clusters of nodes, managing them individually and optimizing intra-cluster communication. This improves scalability by reducing the overall complexity of routing and resource management in large-scale networks.
- **Dynamic Network Slicing:** The SDN controller can enable network slicing, partitioning the network logically to allocate different resources to different types of traffic or services. This allows the network to handle multiple applications (e.g., voice, video, IoT) more efficiently, enhancing scalability without compromising performance.

$$\text{Capacity}_{total} = \text{Capacity}_{clustered} + \text{Capacity}_{sliced}$$

The sum of clustering and network slicing enhances the overall network capacity beyond what is achievable in traditional MANETs. By abstracting certain parts of the network and creating virtual networks, SDN controllers effectively manage larger numbers of nodes without overwhelming the network's resources.

## VI. PERFORMANCE COMPARISON: MANET VS MANET+SDN

To evaluate the performance benefits of integrating SDN into a MANET, we compare the performance of a pure MANET with that of an SDN-enabled MANET using key performance metrics such as latency, throughput, Packet Delivery Ratio (PDR), and control overhead. The comparison is based on theoretical models and available simulation results from studies on SDN-MANET integration [7].

*A. Latency*

In traditional MANETs, latency is often higher due to the distributed nature of routing and the frequent need for route discovery when nodes move. Every time a route is broken, the route discovery process adds significant delay, especially in larger networks.

In an SDN-enabled MANET, the SDN controller maintains a global view of the network and proactively computes optimal routes. This reduces the need for reactive route discovery, lowering the overall latency. The latency in a pure MANET can be modeled as:

$$L_{\text{MANET}} = L_{route\ discovery} + L_{data\ transmission}$$

Whereas, in an SDN-enabled MANET, the latency is primarily dependent on the controller's ability to recompute paths:

$$L_{\text{SDN}} = L_{controller\ computation} + L_{data\ transmission}$$

Since $L_{controller\ computation}$ is typically lower than $L_{route\ discovery}$, SDN-MANETs achieve lower overall latency [5].

*B. Throughput*

Throughput is a measure of the amount of data successfully transmitted over the network. In traditional MANETs, throughput decreases as the number of nodes increases due to higher network congestion and suboptimal routing.

In contrast, SDN controllers can dynamically manage traffic flows and ensure that network resources are used efficiently. By optimizing routing and avoiding congested areas, SDN-enabled MANETs can achieve higher throughput. The total throughput can be expressed as:

$$T_{\text{SDN}} > T_{\text{MANET}}$$

where $T_{\text{SDN}}$ represents the throughput in an SDN-enabled MANET, and $T_{\text{MANET}}$ represents the throughput in a traditional MANET.

*C. Packet Delivery Ratio (PDR)*

The Packet Delivery Ratio (PDR) is the ratio of packets successfully delivered to the destination compared to the number of packets sent. In traditional MANETs, frequent route failures and the need for route rediscovery result in packet losses, reducing the PDR.

SDN-enabled MANETs can maintain higher PDR by quickly adapting to changes in network topology and rerouting packets through alternative paths when links break. Thus:

$$PDR_{SDN} > PDR_{MANET}$$

The improvement in PDR is due to the SDN controller's ability to quickly compute alternative routes and manage traffic flow more efficiently.

*D. Control Overhead*

Control overhead refers to the bandwidth consumed by control messages needed for network management. In traditional MANETs, a significant portion of the available bandwidth is used for control messages, such as route discovery and maintenance in protocols like AODV or OLSR. The control overhead increases significantly in large, dense, or highly mobile networks.

In SDN-enabled MANETs, the centralized controller reduces the number of control messages required for routing decisions, as route computation is handled centrally rather than distributed across nodes. Thus, SDN-enabled MANETs typically have lower control overhead, which can be expressed as:

$$O_{SDN} < O_{MANET}$$

where $O_{SDN}$ is the control overhead in an SDN-enabled MANET, and $O_{MANET}$ is the overhead in a traditional MANET. This reduction in overhead leaves more bandwidth available for data transmission, further improving network performance.

*E. Summary of Performance Metrics*

Table 2 summarizes the expected performance improvements in SDN-enabled MANETs compared to traditional MANETs.

Table 2. Performance comparison: MANET vs SDN-MANET

| Metric | MANET | SDN-MANET |
|---|---|---|
| Latency | High | Lower |
| Throughput | Lower | Higher |
| Packet Delivery Ratio (PDR) | Lower | Higher |
| Control Overhead | High | Lower |

From this comparison, it is clear that integrating SDN into MANETs significantly improves network performance across all key metrics.

## VII. CONCLUSION

Integrating Software-Defined Networking (SDN) with Mobile Ad Hoc Networks (MANETs) provides significant advantages over traditional MANET architectures, particularly in terms of scalability, cost efficiency, and security. The centralized control plane introduced by SDN enables more efficient routing, dynamic resource allocation, and better management of node mobility, which are key challenges in traditional MANETs.

The CAPEX model demonstrates that SDN-enabled MANETs can reduce overall capital expenditures by relying on general-purpose hardware at individual nodes while shifting the computational burden to the SDN controller. In traditional MANETs, every node must independently handle routing and network management, resulting in higher hardware and software costs. By offloading these tasks to a centralized SDN controller, SDN-enabled MANETs reduce the per-node cost and achieve greater scalability as the number of nodes increases.

The OPEX model similarly shows that SDN-enabled MANETs incur lower operational costs by centralizing the maintenance, configuration, and monitoring tasks. Traditional MANETs suffer from high operational costs due to the decentralized nature of the network, where each node must be independently configured and maintained. SDN-enabled MANETs reduce this burden by automating many of these tasks through the SDN controller, improving the network's operational efficiency.

In terms of network efficiency, the global view maintained by the SDN controller allows for optimized routing and dynamic load balancing, leading to better bandwidth utilization and lower latency. Traditional MANETs, which rely on local information for routing decisions, often experience inefficiencies due to suboptimal path selection and delayed route recovery.

The scalability model highlights the potential for SDN-enabled MANETs to handle larger networks more effectively. The SDN controller's ability to manage resources centrally and dynamically adapt to changes in network topology ensures that the network remains scalable even as the number of nodes grows. This contrasts with traditional MANETs, where the lack of centralized control often leads to network congestion and degraded performance in large networks.

Future Work: While this paper presents a theoretical comparison of SDN-enabled MANETs and traditional MANETs, future work should focus on real-world implementations and case studies to validate the proposed models. Research could explore the deployment of SDN-enabled MANETs in environments such as disaster recovery, military operations, and smart cities, where network scalability and flexibility are critical. Additionally, further optimization of SDN controllers for resource-constrained environments, such as those found in MANETs, could be explored to improve the overall performance of these networks.

In conclusion, SDN-enabled MANETs offer a compelling solution for the next generation of mobile ad hoc networks, with substantial improvements in scalability, cost efficiency, and network management. As SDN technology continues to evolve, its integration with MANETs will become increasingly viable, providing a robust framework for managing dynamic, decentralized networks in a wide range of applications.